\documentclass{cdclass}

\usepackage{hyperref}
\usepackage{enumerate}

\usepackage{amsmath}
\usepackage{dsfont}
\usepackage{amsfonts}

\begin{document}

\title{Towards modelling abreast to in-file\\ transitions in small pedestrian groups}
\titlerunning{Towards modelling abreast to in-file transitions in small pedestrian groups}

\author{
  Alessandro Corbetta\authorlabel{1,2} \and
  Chiel van der Laan\authorlabel{1,2} \and
  Francesco Zanlungo\authorlabel{3}
}
\authorrunning{A. Corbetta \and C. van der Laan \and F. Zanlungo}
\institute{
  \authorlabel{1} Department of Applied Physics and Science Education,
  \authortab{1} Eindhoven University of Technology, Eindhoven, The Netherlands,
  \authoremail{1}{a.corbetta@tue.nl}
  \and
  \authorlabel{2} Eindhoven Artificial Intelligence Systems Institute,
  \authortab{2} Eindhoven, The Netherlands
  \and
  \authorlabel{3} Department of Physics and Chemistry ``Emilio Segr\`e'',
  \authortab{3} University of Palermo, Palermo, Italy,
  \authoremail{3}{francesco.zanlungo@unipa.it}
}

\maketitle

% \linenumbers
\begin{abstract}
Pedestrians in a crowd largely walk in small groups, dyads above all, and
modelling how these groups arrange themselves is a step towards realistic
models of the crowd as a whole. A dyad walks side by side in a dilute crowd
and one behind the other in a dense one, an abreast to in-file transition
recently measured in real life across the full range of naturally occurring
crowd conditions (van der Laan and Corbetta, PLOS ONE 2026). The cost-function models of dyad formation by Zanlungo and coworkers (Phys. Rev. E
2014; EPL 2015) hold the pair abreast and have no in-file minimum at any finite
density. Building on them, we propose and analyse a qualitative model in which
the transition appears as an effective density parameter grows: the two
formations become minima that coexist over a window of densities and exchange
stability within it. We quantify this transition by computing the Orientation
Log-Odds $\Pi_\text{OLO}$ as predicted by the model. This
paper is a first step towards quantitative models of dyads and of their
formation transitions.
\end{abstract}


\section{Introduction}

Pedestrian crowds are not collections of independent individuals: a
substantial fraction of pedestrians walk in socially connected groups, and the
groups shape the speed, the spacing and the spatial organisation of the crowd
around them \cite{Moussaid2010,Z2017,Nicolas2021}. Dyads are the most frequent
of these groups \cite{Moussaid2010,Nicolas2021} and the minimal setting in which social cohesion, locomotion
and crowd-induced constraints interact. Their formation is not fixed: a dyad
walks abreast in a dilute crowd and in-file, one member behind the other, in a
dense one (a formation also called ``river-like''
\cite{karamouzas2011simulating}), a transition that some of the authors have recently measured across
the full range of naturally occurring crowd conditions \cite{vdLC2026}.
Modelling the behaviour of groups is a key step towards more realistic models
of the crowd as a whole. Rule-based models have proposed transitions between group formations, e.g.
between abreast and in-file formations in small groups
\cite{karamouzas2011simulating}, as a way to avoid collisions while keeping
the group together. A theoretical study and modelling of such transitions is,
however, still missing, naturally so, since data resolving them have become
available only now. Yet a model for dyads is expected to display a bifurcation
as the crowd conditions that promote the in-file formation set in. Here, adopting the cost-function
perspective previously proposed by some of the authors \cite{Z2014,Z2015EPL},
we establish modelling guidelines for this transition and make a first
qualitative proposal.

The influence of groups on the crowd is well documented: they move more slowly
than isolated pedestrians, occupy more of the transverse space and rearrange
in response to the surrounding flow \cite{Nicolas2021}, and larger groups
organise into
relatively stable dyadic and triadic subunits, which makes dyads the natural
starting point for group-aware models of pedestrian dynamics
\cite{Z2013,Z2014}. Their empirical study has recently changed scale.
Anonymous sensing and automated tracking now yield continuous, months-long
observations of real-life pedestrian traffic
\cite{Gabbana2022,Pouw2024,vdLC2026}, complementary to the
controlled and repeatable conditions of laboratory campaigns
\cite{Seyfried2005,Boltes2013}, and the resulting ensembles of millions of trajectories
allow a statistical description of the inherently disordered dynamics of
crowds: not only mean behaviour, but fluctuations, rare configurations and
conditional dependencies on density, speed and relative flow.

In a dilute crowd dyads walk abreast, the two members side by side at a
preferred interpersonal distance, a length scale we refer to as $r_0$
henceforth, which is compatible with conversation and visual contact
\cite{Moussaid2010,Z2014}; distance and alignment depend on intrinsic features
of the pair such as social relation, walking purpose, gender, age and body
size \cite{Z2017}. In denser environments the pair increasingly walks in-file,
which reduces its transverse footprint and may ease the passage through
restricted regions. The measurements of \cite{vdLC2026} provide a
high-resolution picture of this transition, resolving, among others, the
relative position of the dyad members, and they establish two facts that
constrain any model. First, the transition depends not only on density but on
the motion of the crowd relative to the pair: it is enhanced in standing and
counter-flowing crowds and weaker in co-flowing ones. Second, the
distributions of relative positions peak at the abreast and at the in-file
formations, and the dyad essentially never finds itself in between; hence, the transition is better
looked at as a competition between two distinct states than as a preferred
orientation rotating continuously from one to the other. The Orientation
Log-Odds $\Pi_\text{OLO}$ introduced in \cite{vdLC2026}, the logarithm of the
ratio between the probabilities of the two formations, is the compact measure
of that competition: it is positive where the abreast formation is the more
probable one, negative where the in-file one is, and the transition is where
it changes sign.

Several modelling approaches have been proposed for group features.
Rule-based and force-based models combine individual goal-directed motion,
collision avoidance and attractive or cohesive interactions between group
members \cite{Helbing1995,Moussaid2010,karamouzas2011simulating,Crociani2018}. The cost-function approach we adopt
was introduced in \cite{Z2014}: a radial contribution
favouring a preferred interpersonal distance and an angular contribution
favouring configurations in which the members remain within one another's
visual field, which within a Boltzmann-like description determine a
probability distribution of relative positions and reproduce the
characteristic dyad and triad formations. Subsequent work extended it to
intrinsic group properties \cite{Z2017} and to environmental effects,
including the influence of crowd density on group geometry
\cite{Z2015EPL}, a model calibrated on the group data of \cite{Z2015PRE}. It is the natural starting
point here because it ties an explicit landscape in relative-coordinate space
to a distribution, which the stochastic character of the observations
requires.

As it stands, however, the framework of \cite{Z2014,Z2015EPL} cannot describe
the transition. The
existing cost functions have a single preferred formation,
the abreast one, which the crowd can tilt but never replace: as the density
grows the preferred angle drifts away from abreast towards in-file without
ever reaching it, and no choice of parameters changes this. The measurements,
on the other hand, show two formations, each preferred in its own range of
conditions, which swap roles in between. A cost function for the transition
must therefore hold both formations as minima and let the crowd exchange their
stability, i.e. it must bifurcate, and it must do so under a crowd parameter
that is not the density alone.

In this paper we take a bifurcation between the abreast and the in-file
minima, driven by an effective crowd parameter, as the guideline to propose a
first qualitative cost-based model for the formation transition, for dyads
crossing a standing crowd, the simplest of the flow regimes. The model consists
of two modifications of \cite{Z2015EPL}: a transverse penalty linear rather
than quadratic in the lateral extent of the pair and a softening of the
angular cost. Sampled in a Boltzmann-like fashion, our model reproduces the
abreast to in-file transition, which we quantify in terms of the Orientation
Log-Odds $\Pi_\text{OLO}$ \cite{vdLC2026}. Our contribution provides a first
step focusing on qualitative aspects, and we leave quantitatively accurate
models to future work, as discussed in \sref{sec:discussion}.

This paper is structured as follows. In \sref{sec:frame} we define the
frame of reference we use to analyse and model dyads. \sref{sec:data}
reviews the experimental campaign of \cite{vdLC2026}, focusing on dyads
walking through a standing crowd. \sref{sec:earlier} reports on earlier
cost-function models for dyad formation \cite{Z2014,Z2015EPL}. Our model is
proposed and analysed in \sref{sec:model}. The discussion in
\sref{sec:discussion} closes the paper.

\section{The dyad frame and its reduced variables}
\label{sec:frame}

\begin{figure}[t]
  \begin{center}
    \includegraphics[width=0.55\textwidth]{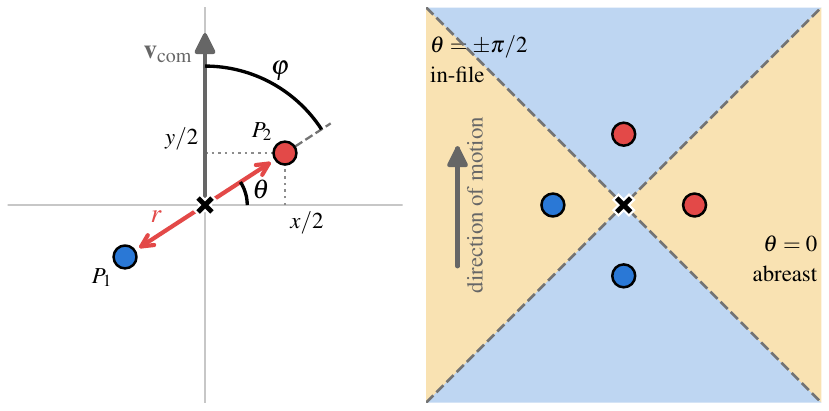}
    \caption{The frame of the dyad model. Left: the origin is the centre of
      mass, $P_1$ and $P_2$ are at $\mp\mathbf{r}/2$, $\theta$ is the angle of
      their separation $\mathbf{r}$ from the abreast axis and $r=\mathbf{r}$ is the magnitude of such separation, $\varphi$ the
      dyad angle, i.e. the angle of the separation from the direction of
      motion (so that $\theta=\pi/2-|\varphi|$), and $u=r/r_0$
      the separation measured in units of the preferred abreast distance; $x$ is
      transverse to the motion and $y$ along it. Right: the abreast and in-file
      configurations, and the partition of the plane into the cones
      $\Gamma^{\leftrightarrow}$ (amber) and $\Gamma^{\updownarrow}$ (blue) of
      \sref{sec:olo}.}
    \label{fig:sec2-frame}
    \end{center}
\end{figure}

We call the two pedestrians in a dyad $P_1$ and $P_2$, and work in a reference frame
centred on their centre of mass, with the $y$ axis aligned with the centre of
mass velocity, as in \fref{fig:dyad-example} and in \cite{vdLC2026}. The
distance vector between the two is
\begin{equation}
\label{erel}
\mathbf{r}=P_2-P_1=\begin{pmatrix}x&y\end{pmatrix},
\end{equation}
so that $P_1$ and $P_2$ are located, respectively, at $\mp\mathbf{r}/2$, and
exchanging the two members is the reflection $\mathbf{r}\to-\mathbf{r}$, a
symmetry that all the quantities below respect. In polar variables we write
\begin{equation}
\label{epolar}
r=\sqrt{x^2+y^2},
\qquad
\theta=\arctan{\frac{y}{|x|}} .
\end{equation}
The absolute value makes $\theta$ blind to which member walks on the left,
at the price of making the quantities that depend on it, such as the costs,
not differentiable at $x=0$: there they have a kink, i.e. different slopes on
the two sides. That axis is
where the in-file configuration lies, so its stability cannot be settled by
setting a derivative to zero; \sref{sec:model} tests it as an endpoint
instead. We measure separations in units of the preferred abreast distance
$r_0$, i.e. throughout the formulas below we use the normalised distance
\begin{equation}
\label{eu}
u=\frac{r}{r_0} .
\end{equation}
In these variables $\theta=0$ identifies the abreast configuration and
$\theta=\pm\pi/2$ the in-file one, cf. \fref{fig:sec2-frame}. The measurements
of \cite{vdLC2026} and of \sref{sec:data} read the formation off the dyad
angle $\varphi$ instead, i.e. the angle between the relative position of the
two members and the direction of motion of their centre of mass,
$\mathbf{v}_\text{com}$, so that $\theta=\pi/2-|\varphi|$:
$\varphi=\pm\pi/2$ is the abreast configuration and $\varphi=0,\pi$ the
in-file one (cf. \fref{fig:sec2-frame}; \fref{fig:dyad-example} reports
$\varphi$ along a dyad trajectory).

\section{Dyads through different crowd densities and the in-file transition}
\label{sec:data}

\subsection{Real-life dyad analysis at Eindhoven Central Station}

\begin{figure}[t]
  \begin{center}
    \includegraphics[width=0.86\textwidth]{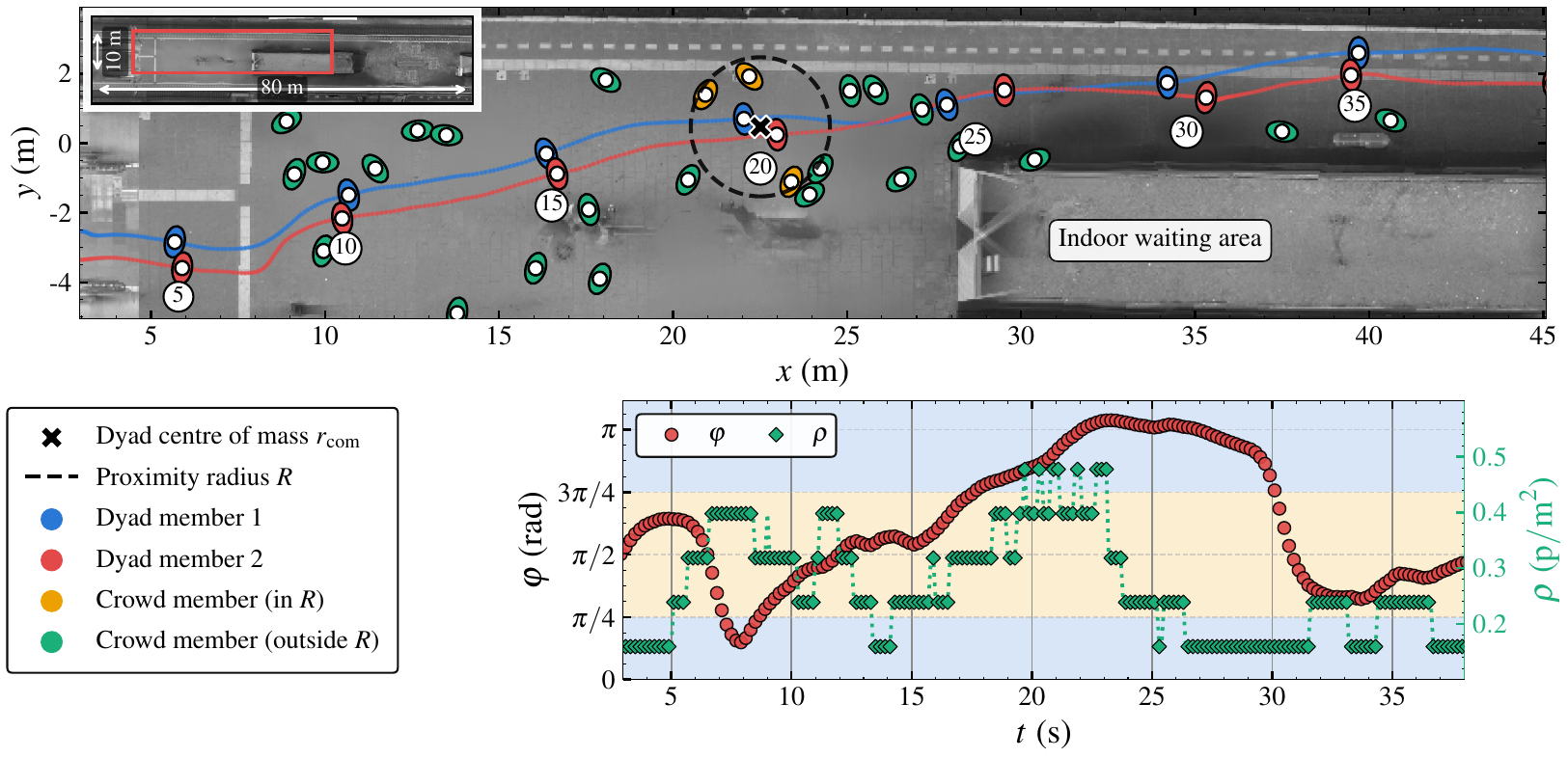}
    \caption{An example dyad crossing the measurement area, replotted from
      Fig. 1(a,d) of \cite{vdLC2026} with a different colour scheme. Top: a section
      of platform 3-4 at Eindhoven Central Station. The two dyad members (blue
      and red) enter from the bottom left and leave at the top right, with
      markers every 5 s labelled by time. The remaining pedestrians are drawn at
      the single snapshot $t=20$ s, and are coloured according to whether they
      fall inside (amber) or outside (green) the dashed proximity disc of
      radius $R=2$ m centred on the dyad centre of mass, which is what defines
      $\rho$ and $\mathbf{v}_\text{prox}$. The inset locates the crop within the
      $80\times 10$ m$^2$ tracked area. Bottom: dyad angle $\varphi$ (red) and
      local density $\rho$ (green) along the same passage, with the abreast
      (amber) and in-file (blue) sectors shaded, and the snapshots of the upper
      panel marked by vertical lines. Around $t=20$ s the dyad enters a denser
      region and turns in-file, returning abreast once the density drops.}
    \label{fig:dyad-example}
    \end{center}
\end{figure}

The data we consider henceforth come from a real-life measurement campaign at Eindhoven Central Station, introduced by some of the authors in \cite{Pouw2024} and partially publicly available \cite{Pouw2024data}. A grid of overhead sensors covers an $80\times 10$ m$^2$ section of platform 3-4 and tracks, anonymously and 24/7, every pedestrian crossing it at 10\,Hz; over the 900 days between March 2021 and March 2024 this amounts to about 21\,M trajectories, roughly 24\,k per weekday, at densities that typically range from free flow up to about 1.3 p/m$^2$ \cite{vdLC2026}. Since the data are anonymous, groups are not annotated and have to be identified a posteriori from the trajectories alone, on the basis of proxemics and of the time consistency of the mutual distance, which yields an ensemble of about 6\,M dyads; see \cite{vdLC2026} for the full detection pipeline.

\fref{fig:dyad-example} reports one such dyad, together with the observables used throughout. The state of the crowd around a pair is summarised through the pedestrians within a disc $\mathcal{D}_R$ of radius $R=2$ m centred on the dyad's centre of mass. Following \cite{vdLC2026}, the local crowd state is defined in terms of the local density
\begin{equation}
\label{erho}
\rho=\frac{N_R}{\pi R^2},
\end{equation}
with $N_R$ the number of pedestrians in $\mathcal{D}_R$, the dyad included (which gives $\rho_\text{free}=0.16$ p/m$^2$ for a dyad walking alone), and of the proxy crowd velocity
\begin{equation}
\label{evprox}
\mathbf{v}_\text{prox}=\frac{1}{N_R}\sum_{i\in\mathcal{D}_R}\mathbf{v}_i ,
\end{equation}
i.e. the mean velocity of the pedestrians in $\mathcal{D}_R$. The magnitude of $\mathbf{v}_\text{prox}$ defines the standing crowd regime of \sref{sec:standing}. The formation is read off the dyad angle $\varphi$ of \sref{sec:frame}.

\begin{figure}[t]
  \begin{center}
    \includegraphics[width=0.88\textwidth]{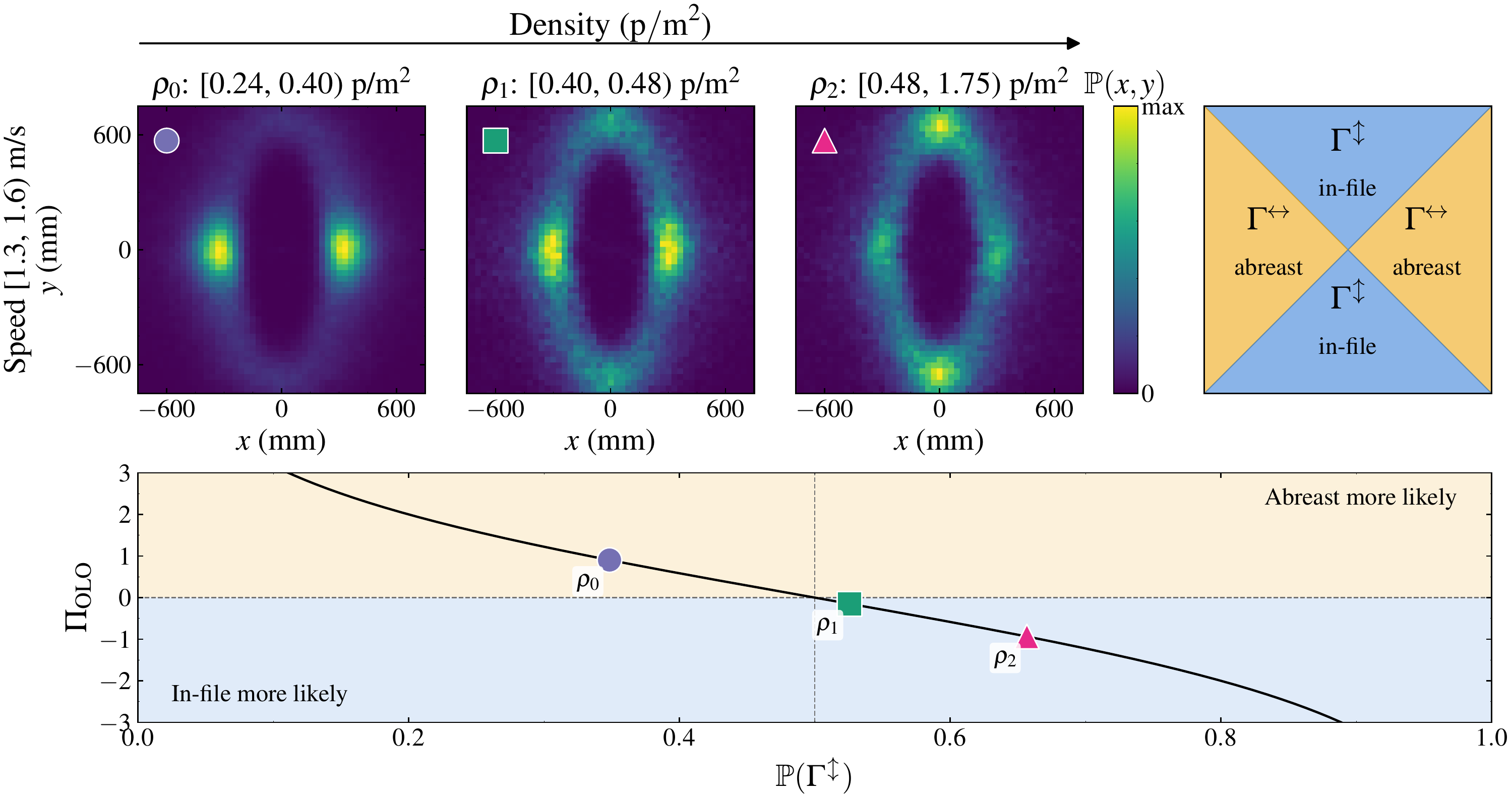}
    \caption{From the distribution of relative positions to $\Pi_\text{OLO}$,
      for dyads walking through a standing crowd at various density levels
      $\rho$, replotted from \cite{vdLC2026}. Top: the relative position
      distribution function, $\mathbb{P}(x,y)$, in three
      bins of increasing density, and the partition of \eref{eqcones} into
      the abreast cones $\Gamma^{\leftrightarrow}$ (amber) and the in-file
      cones $\Gamma^{\updownarrow}$ (blue). Each panel is scaled
      with respect to its maximum probability density level for visualisation
      purposes. Bottom: the map from $\mathbb{P}(\Gamma^{\updownarrow})$ to
      $\Pi_\text{OLO}$ (\eref{eqolo}); the markers report the
      $\Pi_\text{OLO}$ values for the $\mathbb{P}(x,y)$ depicted.}
    \label{fig:olo-curve}
    \end{center}
\end{figure}

One observation is decisive for what follows. Whatever the density, the maxima of the distribution of relative positions are at the abreast or at the in-file formation, $\theta=0$ or $\theta=\pm\pi/2$, and essentially never at intermediate angles (\fref{fig:olo-curve}).

\subsection{$\Pi_\text{OLO}$ as a synthetic indicator}
\label{sec:olo}

Analysing the full probability density distribution $\mathbb{P}(x,y)$ of the relative displacement of \eref{erel} across crowd conditions is impractical: every condition yields a two dimensional field, while the feature we want to track is a single qualitative one, i.e. whether the dyad walks abreast or in-file. In \cite{vdLC2026} the field is therefore compressed into one scalar, the Orientation Log-Odds $\Pi_\text{OLO}$. To compute $\Pi_\text{OLO}$, the $(x,y)$ plane is split by its two diagonals into two pairs of opposite cones of equal area,
\begin{equation}
\label{eqcones}
\Gamma^{\leftrightarrow}=\left\{|y|\leq|x|\right\},\qquad
\Gamma^{\updownarrow}=\left\{|y|>|x|\right\},
\end{equation}
so that $\Gamma^{\leftrightarrow}$ collects the abreast configurations and $\Gamma^{\updownarrow}$ the in-file ones. The Orientation Log-Odds is then defined as
\begin{equation}
\label{eqolo}
\Pi_\text{OLO}=\log_2\frac{\mathbb{P}(\Gamma^{\leftrightarrow})}{\mathbb{P}(\Gamma^{\updownarrow})}.
\end{equation}
$\Gamma^{\leftrightarrow}$ and $\Gamma^{\updownarrow}$ have equal area, hence a dyad with no preference gives $\Pi_\text{OLO}=0$; positive values say that the abreast state is the more likely one, negative values that the in-file state is, and the magnitude counts the factors of two separating the two likelihoods: e.g. $\Pi_\text{OLO}=-1$ means that the in-file state is twice as likely as the abreast one.

$\Pi_\text{OLO}$ is a logit and, notably, under a Boltzmann-like assumption (\eref{eqboltz}) it is, up to a temperature, a difference in free energy between the abreast and the in-file formation.

\fref{fig:olo-curve} shows $\Pi_\text{OLO}$ on measured data in the standing-crowd regime (\sref{sec:standing}) for dyads with $v_\text{com}\approx1.4$~m/s, across three bins of increasing density $\rho_0<\rho_1<\rho_2$. The maxima of the distribution move with the density: at $\rho_0$ they are on the transverse axis, at $\rho_1$ the distribution is close to a ring and carries both an abreast and an in-file pair of maxima, and at $\rho_2$ the in-file pair dominates, though fainter abreast local maxima persist in the ring. The three panels give, respectively, $\Pi_\text{OLO}=+0.83$, $-0.22$ and $-1.01$. Notably, the typical distance between the dyad members differs between the abreast and the in-file configuration, a feature that we here neglect by considering a single preferred distance scale $r_0$ (see \sref{sec:discussion}).

\begin{figure}[t]
  \begin{center}
    \includegraphics[width=0.88\textwidth]{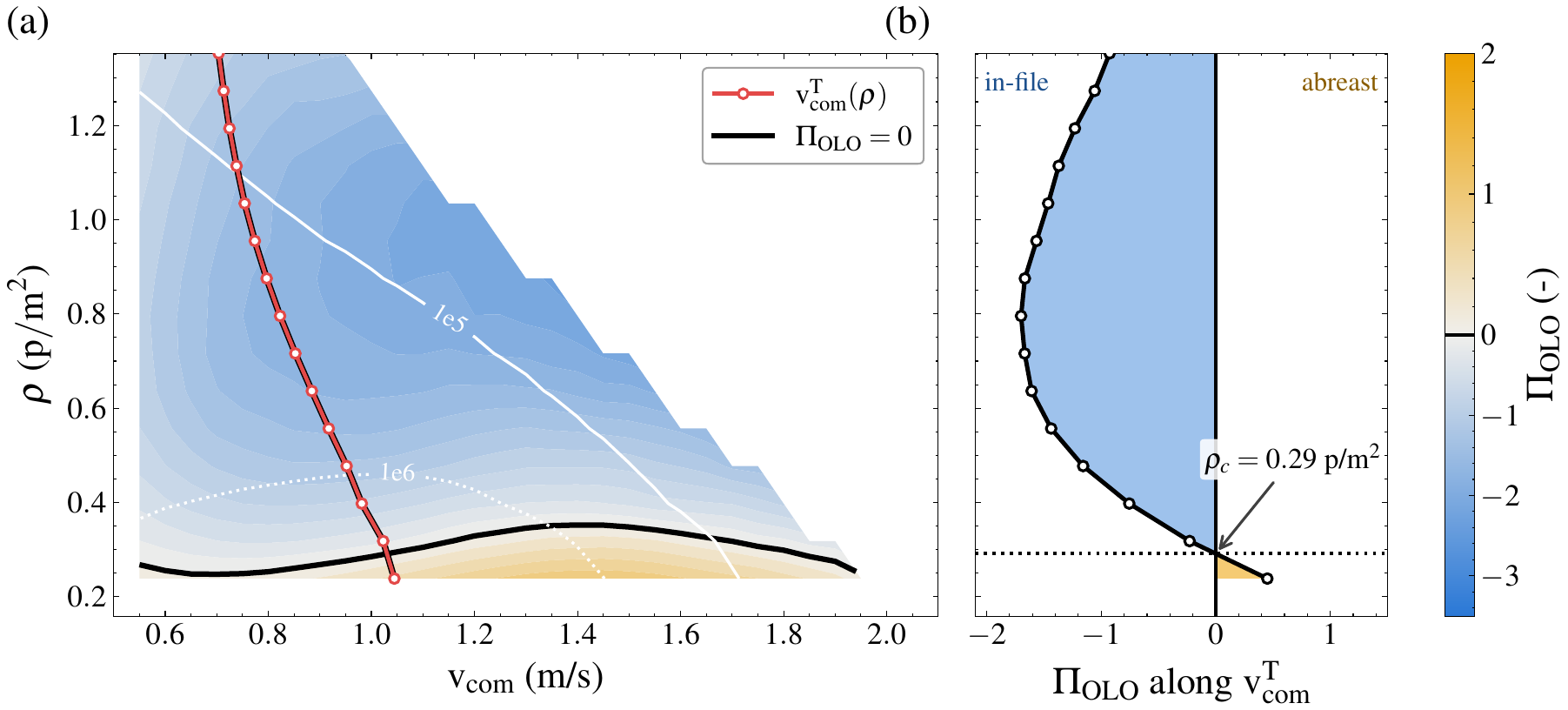}
    \caption{Measured $\Pi_\text{OLO}$ for a dyad traversing a standing crowd,
      replotted from \cite{vdLC2026}. (a) The surface over the
      $(v_\text{com},\rho)$ plane. Amber is abreast dominant
      ($\Pi_\text{OLO}>0$), blue in-file dominant, as everywhere else in this
      paper; the thick black line is the $\Pi_\text{OLO}=0$ transition and the
      white contours count the measurements supporting each region. The red
      trace is the typical speed $v_\text{com}^\text{T}(\rho)$, the
      count-weighted mean speed at each density, as in the fundamental diagram
      of \cite{vdLC2026}. (b) $\Pi_\text{OLO}$ measured on the typical
      velocity-density state for a dyad in a standing crowd. The sign changes
      at $\rho_c\simeq0.29$ p/m$^2$, while being in general a non-monotone
      relation.}
    \label{fig:olo-map}
    \end{center}
\end{figure}

\subsection{Dyads in a standing crowd}
\label{sec:standing}

We restrict now to the standing crowd regime, here defined as the conditions in which the crowd around a dyad moves below the threshold $v_s=0.4$ m/s, i.e. $|\mathbf{v}_\text{prox}|<v_s$. This is the simplest of the flow regimes, in that the crowd enters through its density alone and not through a relative motion (cf. \cite{vdLC2026} for analyses beyond the standing case).

\fref{fig:olo-map} reports the measured $\Pi_\text{OLO}$ over the $(v_\text{com},\rho)$ plane in this regime, as in \cite{vdLC2026}. The black line is the $\Pi_\text{OLO}=0$ contour, i.e. the abreast to in-file transition: below it the abreast state is the more likely one, above it the in-file state is. The transition is at a density of about $0.3$ p/m$^2$ and depends only weakly on the speed, being highest, i.e. most favourable to the abreast state, around $v_\text{com}\approx1.4$ m/s. The white contours count the measurements behind each region: high speeds at high density are rare, as expected, the feasible speed through a crowd decreasing with density.

$\Pi_\text{OLO}=\Pi_\text{OLO}(v_\text{com},\rho)$ is in general
non-monotonic in both its variables. At fixed density it has a minimum and a
maximum, which approach each other as the density grows and eventually merge
\cite{vdLC2026}. To analyse the non-monotonicity in $\rho$, in
\fref{fig:olo-map}(b) we report
$\Pi_\text{OLO}^\text{T}=\Pi_\text{OLO}(v_\text{com}^\text{T}(\rho),\rho)$,
with $v_\text{com}^\text{T}$ the average measured velocity at density
$\rho$ (see the fundamental diagrams in \cite{vdLC2026}). Along it
$\Pi_\text{OLO}$ shifts from abreast to in-file dominance at the critical
density $\rho_c\simeq0.29$ p/m$^2$, reaches a minimum around
$\rho\simeq0.8$ p/m$^2$, and rises again at the highest densities, possibly
because dyads there switch formation readily. In the following we propose a
model for the decreasing branch of $\Pi_\text{OLO}^\text{T}$, leaving the
non-monotonicity to future work.

\section{Earlier cost-function models of dyad dynamics}
\label{sec:earlier}

We review here the cost functions for dyad geometry proposed by some of the
authors \cite{Z2014,Z2015EPL}, written in the variables of
\sref{sec:frame} and in the form the analysis of
\sref{sec:model} will use, with the crowd strength denoted throughout
by the effective density $\rho_\text{eff}$ of
\sref{sec:model}.\footnote{The asymmetric interaction of \cite{Z2014} is
dropped, i.e. its parameter $\eta$ is set to zero, and a constant is added so
that the cost is zero at its minima.} The models were calibrated on the group
data of
\cite{Z2014,Z2015PRE}, which cover the low to medium density regime only, up to
$0.3$ p/m$^2$, i.e. up to about the density at which the measurements of
\sref{sec:data} place the transition: those data had possibly no reason to
call for an in-file minimum.

\subsection{The zero density cost function}

For a dyad in an otherwise empty environment \cite{Z2014} proposes the cost
function\footnote{In a complete cost function that assumes the presence of
other terms, e.g. related to energy expenditure or collision avoidance, this
cost function may need to be multiplied by a weight, i.e. in principle it will
be proportional to the one discussed here.}
\begin{equation}
\label{e1}
U(u,\theta)=u+\frac{1}{u}+\frac{A}{2} \theta^2 ,
\end{equation}
which \cite{Z2014} introduces as a potential in equations of motion for the
accelerations of the two members: there the cost enters the dynamics through
its gradient, as a force on each member, and the formations are the equilibria of the resulting
equations of motion. Here we use only that static landscape, and read the
formations off its minima. The radial terms give a single minimum at the preferred distance,
and grow without bound away from it,
\begin{equation}
\label{eradial}
\begin{aligned}
\operatorname{sign}(\partial_u U) &= \operatorname{sign}(u-1) && \text{for every } \theta ,\\
\lim_{u\to 0,+\infty} U &= +\infty , &&
\end{aligned}
\end{equation}
and the angular term, with $A>0$, places it at $\theta=0$. \eref{e1} is not
differentiable at $\theta=\pm\pi/2$, which was immaterial in \cite{Z2014}, where
the model described the dynamics near the equilibria, but is not here: the
in-file configuration lies precisely at that point, and we therefore treat it
in \sref{sec:model} as an endpoint rather than as an interior
stationary point.

To compare a cost function with the measurements, \cite{Z2014,Z2015EPL}
assume that the dyad samples it in a Boltzmann-like fashion,
\begin{equation}
\label{eqboltz}
\mathbb{P}(x,y)\propto\exp\left[-\frac{U(u(x,y),\theta(x,y))}{T}\right],
\end{equation}
with a temperature $T$ that collects what the cost function does not
describe, e.g. the noise of walking and the manoeuvres imposed by the
surrounding crowd. The level sets of $U$ are therefore also the contours of
the distribution of relative positions $\mathbb{P}(x,y)$, and the minima of
$U$ its maxima (\fref{fig:sec2-maps}).

\subsection{The density dependent term}

Over the density range they cover, the data of \cite{Z2014TRP,Z2015PRE} give no
evidence of a qualitative density-dependent change in the structure of dyads.
They do show the distribution of relative positions changing with density,
with a decreasing average separation. The
model of \cite{Z2015EPL} reproduces this by adding to \eref{e1} a penalty on the
squared transverse extent of the pair, $C\rho_\text{eff}x^2/r_0^2$, i.e.
\begin{equation}
\label{e2}
U(u,\theta,\rho_\text{eff})
  =u+\frac{1}{u}+\frac{A}{2} \theta^2
  + C\rho_\text{eff}\,u^2\cos^2\theta .
\end{equation}
In line with the notation of \sref{sec:model} we already denote the
crowd strength by $\rho_\text{eff}$. In \cite{Z2015EPL} it is the physical
density $\rho$; the measurements of \sref{sec:data} show
density not to be the only relevant crowd variable, and in 
\sref{sec:model} we promote $\rho_\text{eff}$ to a dimensionless combination of
density and relative crowd velocity, whose form we leave open.

\begin{figure}[t]
  \begin{center}
    \includegraphics[width=0.68\textwidth]{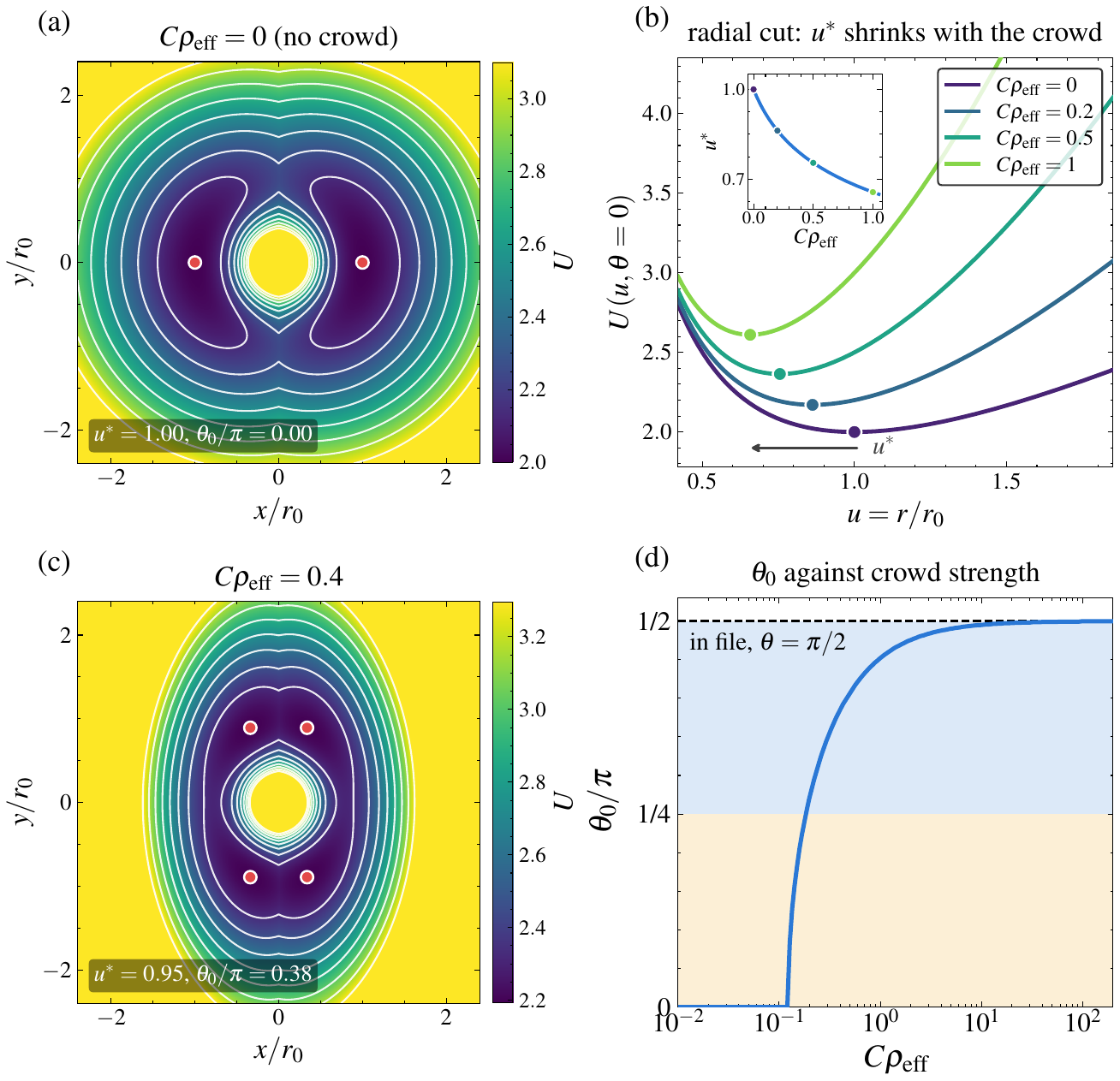}
    \caption{Level sets of the cost functions of \cite{Z2014} and
      \cite{Z2015EPL} in the $(x/r_0,y/r_0)$ plane (cf.
      \fref{fig:sec2-frame}), at $A=0.2$; through \eref{eqboltz} they are
      also the contours of the distribution $\mathbb{P}(x,y)$ the models
      predict. (a) \eref{e1}: closed level
      sets around a single pair of minima on the abreast axis at $u=1$. The
      angular stiffness, $\partial^2_\theta U=A$, is an order of magnitude
      smaller than the radial one, $\partial^2_u U=2$, so the level sets are
      elongated along the circle $u\simeq1$: changing the orientation of the
      dyad costs much less than changing its separation. (b) The cost along
      the abreast axis, $\theta=0$, for increasing crowd strength; inset:
      the position $u^*$ of its minimum at different densities
      $C\rho_\text{eff}$. The transverse penalty is quadratic in $u$, so the
      minimum moves inwards, from $u^*=1$ to $0.66$ at $C\rho_\text{eff}=1$,
      which reproduces the contraction of the average separation reported in
      \cite{Z2015PRE}. (c)
      \eref{e2} at $C\rho_\text{eff}=0.4$: the level sets are elongated
      along the direction of motion and the minima have left the abreast axis.
      (d) The angle $\theta_0$ of the minimum against crowd strength, on a
      logarithmic scale: it leaves $0$ at $C\rho_\text{eff}\simeq0.12$ and
      approaches $\pi/2$ without attaining it, still $0.499\pi$ at
      $C\rho_\text{eff}=2\times10^2$, as according to \eref{ebif-never}.}
    \label{fig:sec2-maps}
    \end{center}
\end{figure}

After calibrating on those data, \cite{Z2015EPL} extrapolates \eref{e2} to
higher densities: the minima remain at $\theta=0$ below a critical density
$\rho^*$ and split into a pair at $\pm\theta_0$ above it, with
\begin{subequations}
\label{ebif}
\begin{align}
\theta_0 &= 0 && \text{for } \rho_\text{eff}<\rho^* , \label{ebif-low}\\
\theta_0 &< \pi/2 && \text{for every } \rho_\text{eff} , \label{ebif-never}\\
\lim_{\rho_\text{eff}\to\infty}\theta_0 &= \pi/2 . && \label{ebif-limit}
\end{align}
\end{subequations}
\fref{fig:sec2-maps} reports the corresponding level sets and the
displacement of the minima. \eref{ebif-never} is the essential
one: $\theta_0$ tends to $\pi/2$ only asymptotically, so the in-file
configuration is not a minimum of \eref{e2} at any finite density\footnote{Although in the high density regime the position of the minimum would be, from an experimental point of view, qualitatively indistinguishable from an in-file one.}, whereas
the measurements of \sref{sec:data} show it to be the dominant one
above a moderate density.

\section{A minimal bifurcating model for dyads' formation}
\label{sec:model}

We look for a cost function whose minima, as a crowd parameter grows, pass
from the abreast to the in-file configuration through an exchange of
stability, i.e. with both configurations being minima over a range of the
parameter, and not through a rotation across intermediate angles, which the
measurements of \sref{sec:data} exclude. We opt to stay as close as possible to the
models of \sref{sec:earlier}.

We operate two simplifications. First, the crowd enters through a single
dimensionless parameter, the effective density $\rho_\text{eff}$. We expect
it to grow with the local density and with the velocity of the crowd relative
to the dyad (\fref{fig:olo-map}); given the non-monotonicity of
$\Pi_\text{OLO}$ with density and velocity, we leave its form open. Second, the abreast and in-file
formations share the same preferred distance $r_0$, although the measured
distributions $\mathbb{P}(x,y)$ have their abreast and their in-file maxima at
different distances (\sref{sec:olo}). \sref{sec:discussion} comes back to
both aspects.

The cost function we propose is
\begin{equation}
\label{eqmodel}
U(u,\theta)=u+\frac{1}{u}
  +\frac{A}{2}\left[\theta^2-\frac{2B}{\pi^2}\theta^4
  +2\rho_\text{eff}\,u\cos\theta\right],
\end{equation}
with $\theta$ and $u$ as in Eqs.~\ref{epolar} and~\ref{eu}. The crowd term,
$A\rho_\text{eff}|x|/r_0$, acts as a repulsive ``charge'' on the lateral extent of the dyad,
which pushes it out of the abreast configuration. The pure number $B$ softens
the angular cost. Besides, $B=1$ is the value at which, at zero density, the slope of
the angular cost at the in-file configuration vanishes, i.e. the in-file
configuration is at the edge of turning from a maximum into a minimum
(\sref{sec:model-corrections}).

\begin{figure}[t]
  \begin{center}
    \includegraphics[width=0.90\textwidth]{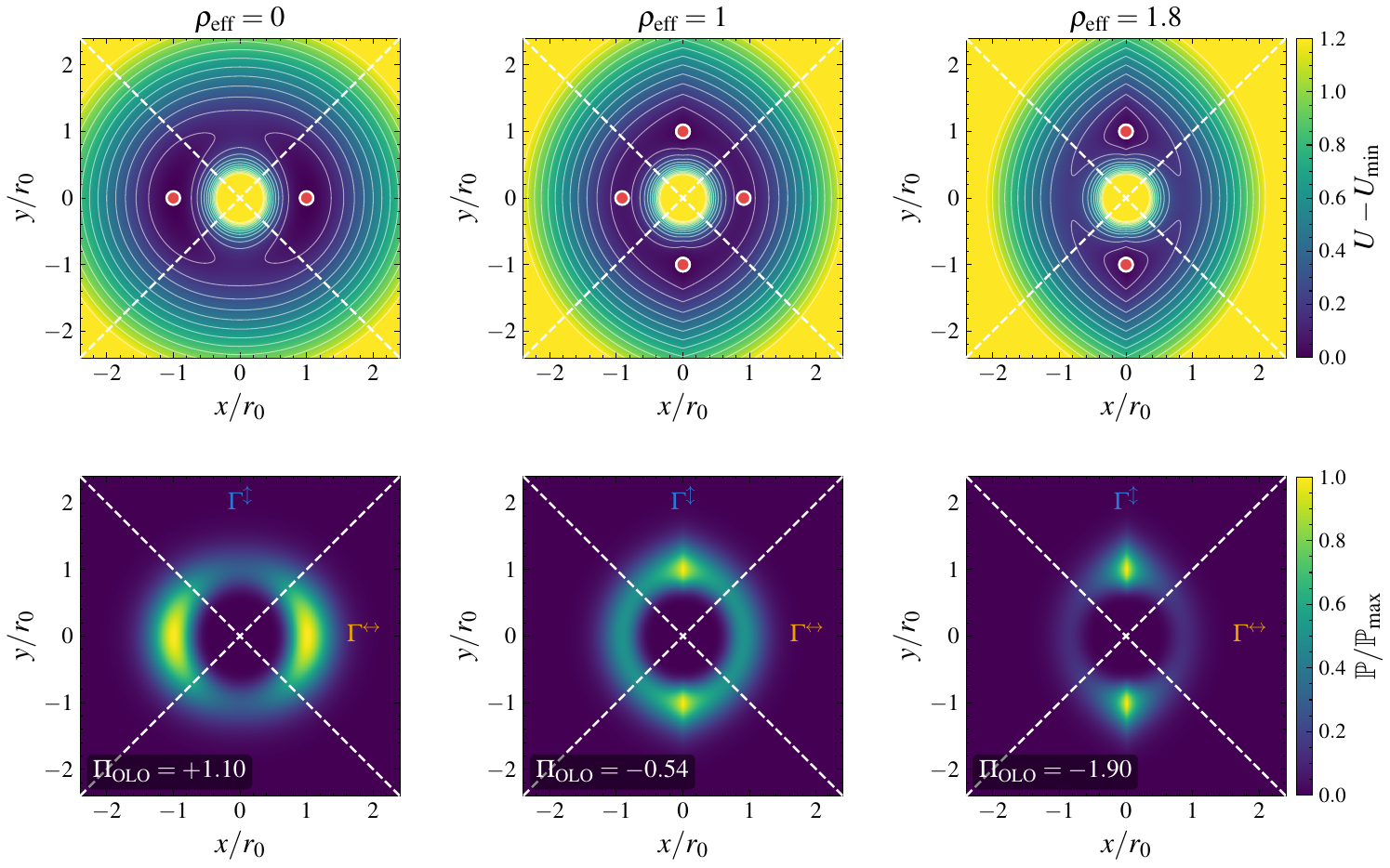}
    \caption{The model of \eref{eqmodel} at $A=0.2$, $B=1$, at
      three densities. Top: the cost $U$ in the $(x/r_0,y/r_0)$ plane of
      \fref{fig:sec2-frame}, measured from its minimum, $U-U_\text{min}$,
      and clipped at $1.2$, so that one colorbar serves each row, with the
      minima of $W$, \eref{eqW}, marked and the diagonals of
      \eref{eqcones} dashed. Bottom: the distribution of \eref{eqboltz} at
      $T=0.1$, each panel normalised by its own maximum as in
      \fref{fig:olo-curve}, with its $\Pi_\text{OLO}$
      (\sref{sec:model-sampling}). Cf.~\fref{fig:olo-curve} for qualitatively
      similar measured distributions.}
    \label{fig:model-landscape}
    \end{center}
\end{figure}

\fref{fig:model-landscape} shows the cost of \eref{eqmodel} at three
densities, together with the distribution it generates
(\sref{sec:model-sampling}): indeed, minima are on the abreast axis in free flow, on both
axes at intermediate density, and on the in-file axis at high density.

\begin{table}[t]
  \begin{center}
    \caption{The symbols of the model, and the values used in the figures.}
    \label{tab:symbols}
    \begin{tabular}{lll}
      \hline
      symbol & meaning & value \\
      \hline
      $u$, $\theta$ & separation in units of $r_0$, angle from the abreast axis & \\
      $A$ & angular stiffness & $0.2$ \\
      $B$ & quartic softening of the angular cost & $B_\text{c}<B\leq1$; $1$ \\
      $\rho_\text{eff}$ & effective density, dimensionless & \\
      $T$ & temperature of the Boltzmann-like sampling & $0.1$ \\
      $W(\theta)$ & cost minimised over the separation, \eref{eqW} & \\
      $\rho_\text{eff}^{\,\text{a}}$, $\rho_\text{eff}^{\,\text{f}}$ & abreast and in-file stability thresholds & $1.10$; $\frac{\pi}{2}(1-B)$ \\
      $B_\text{c}$ & least $B$ for coexistence & $0.30$ \\
      $\rho_\text{eff}^{\,\text{c}}$ & sign change of $\Pi_\text{OLO}$ & $0.67$ \\
      \hline
    \end{tabular}
  \end{center}
\end{table}

\subsection{Improvements over the earlier models}
\label{sec:model-corrections}

\eref{eqmodel} differs from \eref{e2} in two key terms. Both act on the
angular part of the cost, i.e. on what is left of the cost at a fixed
separation $u$ once the radial terms $u+1/u$, common to all the cost functions
of this paper, are removed. Let
\begin{equation}
\label{eqThetadef}
\Theta(\theta;u)\equiv U(u,\theta)-u-\frac{1}{u}
\end{equation}
be that angular part. For the model of \cite{Z2015EPL}, \eref{e2}, it
holds
\begin{equation}
\label{eqtheta0}
\Theta(\theta;u)=\frac{A}{2}\theta^2+\tilde{K}\cos^2\theta,
\qquad
\tilde{K}=C\rho_\text{eff}\,u^2 ,
\end{equation}
and at the in-file configuration we have
\begin{equation}
\label{eqcusp}
\Theta'\left(\frac{\pi}{2}\right)=\frac{A\pi}{2}>0
\end{equation}
at every density. The crowd term, $\tilde{K}\cos^2\theta$, has zero slope
at $\theta=\pi/2$, and thus disappears from \eref{eqcusp}: what is left is
the slope of the social term $A\theta^2/2$, which pulls the pair back towards
the abreast axis. Since this slope is positive, the cost decreases when the
dyad rotates away from the in-file configuration towards the abreast one:
the in-file configuration is a cuspidal maximum, i.e. a maximum at a kink of the cost
rather than at a point of zero slope, at any density, and increasing
$\tilde{K}$ moves the minimum towards $\pi/2$ without reaching it, which is
\eref{ebif-never}.

Our model changes \eref{eqtheta0} in two steps, each addressing one of the
two conditions the data impose: the first makes the in-file configuration a
possible minimum, the second lets it be a minimum together with the abreast
one.

\begin{enumerate}
\item \textbf{Linear transverse penalty.}
Penalising $|x|$ instead of $x^2$ gives
\begin{equation}
\label{eqtheta1}
\Theta(\theta;u)=\frac{A}{2}\theta^2+K\cos\theta,
\qquad K=A\rho_\text{eff}\,u,
\end{equation}
with $\Theta''(0)=A-K$ and $\Theta'(\pi/2)=A\pi/2-K$. The abreast
configuration is a minimum for $K<A$, the in-file one for $K>A\pi/2$; we
call these two values of $K$ the abreast and the in-file thresholds. The
in-file configuration can now be a minimum, but, since $\pi/2>1$, never at the
same density as the abreast one: for $A<K<A\pi/2$, the minimum is at an
intermediate angle $\theta_0$, solution of $A\theta_0=K\sin\theta_0$, which
the data exclude.

\item \textbf{Quartic ``softening'' of the angular cost.} We add the next
even term of the Taylor expansion of the angular cost, in the same spirit as
\eref{e1}, which was proposed as an approximation around the abreast minimum,
\begin{equation}
\label{eqtheta2}
\Theta(\theta;u)=\frac{A}{2}\left[\theta^2-\frac{2B}{\pi^2}\theta^4\right]
  +K\cos\theta .
\end{equation}
The relation $\Theta''(0)=A-K$ holds unchanged with respect to \eref{eqtheta1}, i.e. the
abreast threshold stays at $K=A$, while $\Theta'(\pi/2)=A\pi(1-B)/2-K$, so that the in-file configuration is a
minimum for $K>A\pi(1-B)/2$: the in-file threshold is lowered by the factor
$1-B$. We require $B\leq 1$, since otherwise the in-file configuration is a
minimum already at zero effective density, whereas unconstrained dyads walk abreast. The two
configurations are minima at the same densities, as the data require, if the
in-file threshold falls below the abreast one, $A\pi(1-B)/2<A$, i.e. for
\begin{equation}
\label{eqBfixedu}
B>1-\frac{2}{\pi}\simeq 0.36 .
\end{equation}
Below this value the model behaves as in step 1, with an intermediate minimum
between the two thresholds.
\end{enumerate}

Both modifications are thus needed for the abreast and the in-file
configurations to coexist as minima, and they do so for
$1-2/\pi<B\leq 1$. This estimate holds at a fixed separation $u$;
\sref{sec:model-analysis} lets $u$ take, at each angle, the value that
minimises the cost.

\begin{figure}[t]
  \begin{center}
    \includegraphics[width=0.88\textwidth]{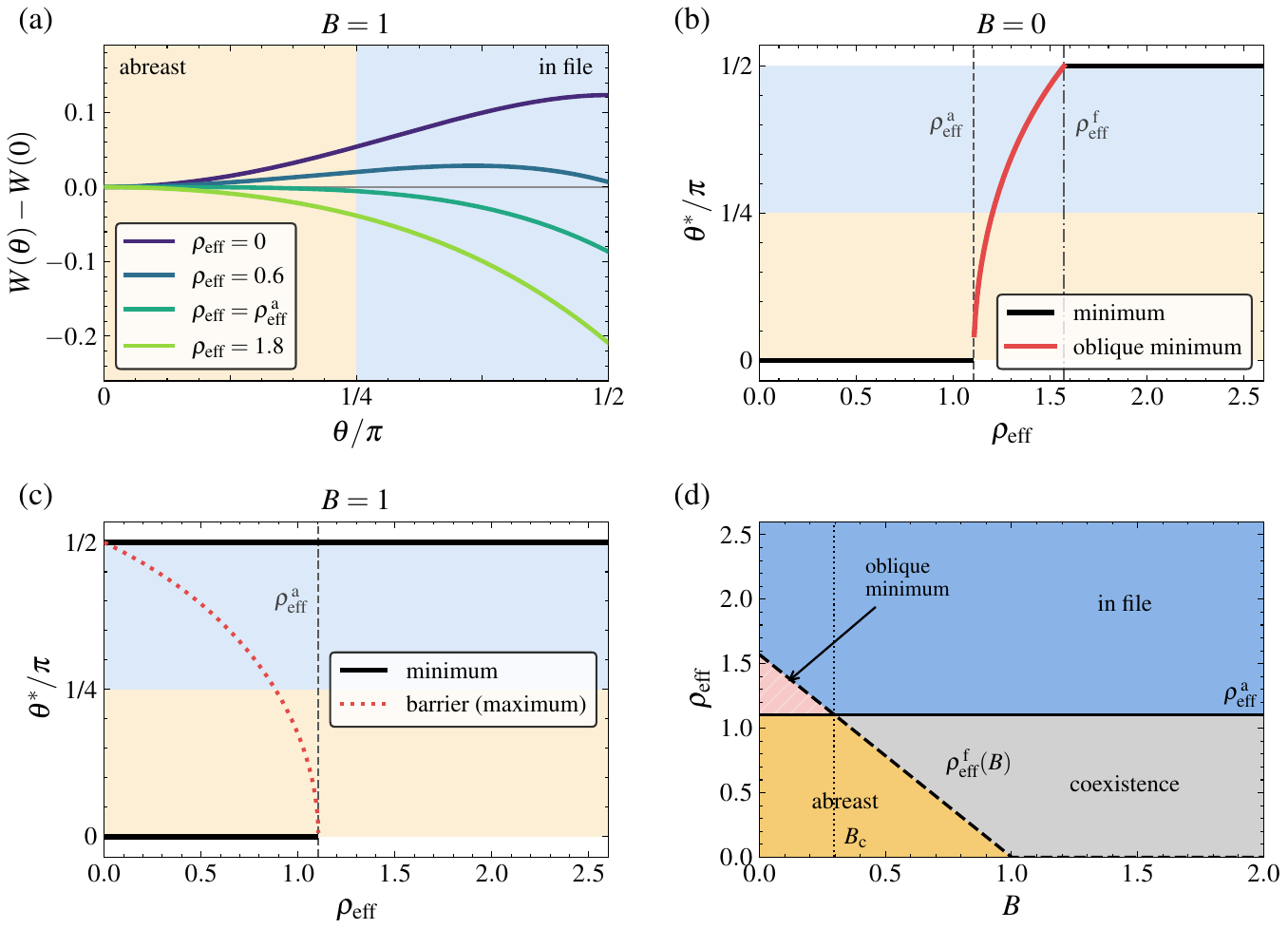}
    \caption{The reduced angular potential $W$ of \eref{eqW} and its
      bifurcation, at $A=0.2$. (a) $W(\theta)-W(0)$ at four densities for
      $B=1$; the abreast ($|\theta|<\pi/4$, amber) and in-file
      ($|\theta|>\pi/4$, blue) cones of \eref{eqcones} are shaded as in
      Figs.~\ref{fig:dyad-example} and~\ref{fig:olo-curve}. (b) The extrema of
      $W$ against $\rho_\text{eff}$ for $B=0$: the abreast minimum is lost at
      $\rho_\text{eff}^{\,\text{a}}$ before the in-file one appears at
      $\rho_\text{eff}^{\,\text{f}}$, so in between the only minimum is the
      oblique one, which rotates continuously from $0$ to $\pi/2$. (c) The
      same for $B=1$: the two configurations coexist up to
      $\rho_\text{eff}^{\,\text{a}}$, and the interior stationary point is a
      maximum, i.e. the barrier between them. (d) Phase diagram in the
      $(B,\rho_\text{eff})$ plane, bounded by Eqs.~\ref{eqrhoa}
      and~\ref{eqrhof}: the oblique region (hatched) is confined to
      $B<B_\text{c}$.}
    \label{fig:model-bifurcation}
    \end{center}
\end{figure}

\subsection{Analysis of the model and its bifurcation}
\label{sec:model-analysis}

In this section we show that the stability of the two configurations can be
worked out exactly also when the separation $u$ is not fixed but takes, at
each angle, the value that minimises the cost, and derive the
densities at which each of them is a minimum. The separation $u$ can be
eliminated from \eref{eqmodel}. In fact, since $\partial_u^2 U>0$, at each
$\theta$ the cost has a single radial minimum,
\begin{equation}
\label{equstar}
u^*(\theta)=\frac{1}{\sqrt{1+A\rho_\text{eff}\cos\theta}} ,
\end{equation}
and inserting it in \eref{eqmodel} gives
\begin{equation}
\label{eqW}
W(\theta)\equiv\min_u U(u,\theta)
  =2\sqrt{1+A\rho_\text{eff}\cos\theta}
   +\frac{A}{2}\left[\theta^2-\frac{2B}{\pi^2}\theta^4\right].
\end{equation}
In words, $W(\theta)$ retains, for each orientation $\theta$, the cost at the
most favourable separation, so that the minima of $U$ in the plane are the
minima of $W$ on $[0,\pi/2]$, at distance $u^*(\theta)$. They can be at $\theta=0$
(abreast), at $\theta=\pi/2$ (in-file), or at an intermediate angle, which we
call \emph{oblique}. The oblique case is the behaviour of \eref{e2} once its
minima have left the abreast axis (\fref{fig:sec2-maps}(c,d)), which the data
exclude. The signs of
$W''(0)$ and of $W'(\pi/2)$ give three conditions.
\begin{itemize}
\item \emph{Abreast.} The abreast configuration is a minimum for
  \begin{equation}
  \label{eqrhoa}
  \rho_\text{eff}<\rho_\text{eff}^{\,\text{a}}
    =\frac{A+\sqrt{A^2+4}}{2} ,
  \end{equation}
  independently of $B$.
\item \emph{In-file.} The in-file configuration is a (cuspidal) minimum for
  \begin{equation}
  \label{eqrhof}
  \rho_\text{eff}>\rho_\text{eff}^{\,\text{f}}(B)
    =\frac{\pi}{2}\left(1-B\right).
  \end{equation}
\item \emph{Coexistence.} Both are minima over a range of densities if
  $\rho_\text{eff}^{\,\text{f}}(B)<\rho_\text{eff}^{\,\text{a}}$, i.e. for
  \begin{equation}
  \label{eqBc}
  B>B_\text{c}=1-\frac{A+\sqrt{A^2+4}}{\pi} .
  \end{equation}
\end{itemize}
For $A\to0$, \eref{eqBc} reduces to the fixed-distance bound of
\eref{eqBfixedu}; letting the distance relax lowers $B_\text{c}$. The
abreast distance shrinks with density as
$u^*(0)=(1+A\rho_\text{eff})^{-1/2}$, and the data do not show a comparable
contraction, so we keep $A$ small: $A=0.2$ in all figures of this section, for
which $u^*(0)$ decreases by 9\% at $\rho_\text{eff}=1$, and
$\rho_\text{eff}^{\,\text{a}}=1.10$, $B_\text{c}=0.30$.

\fref{fig:model-bifurcation} shows the two regimes. For $B<B_\text{c}$,
panel (b), the abreast minimum is lost at $\rho_\text{eff}^{\,\text{a}}$
before the in-file one appears at $\rho_\text{eff}^{\,\text{f}}$, and in
between the only minimum is oblique and rotates continuously from $0$ to
$\pi/2$, as for \eref{e2}. For
$B>B_\text{c}$, panel (c), the interior stationary point is a maximum, i.e.
the barrier between two stable configurations, and the dyad is either abreast
or in-file. In Panel (d) we report the complete phase diagram.

\subsection{The distribution and the $\Pi_\text{OLO}$ of the model}
\label{sec:model-sampling}

As in \cite{Z2014,Z2015EPL}, we sample the cost function of \eref{eqmodel}
through the Boltzmann-like weight of \eref{eqboltz}, treating the temperature
$T$ as a free parameter.

The lower row of \fref{fig:model-landscape} shows the resulting distribution
at $B=1$. The progression is qualitatively similar to the measured one of \fref{fig:olo-curve}: two
abreast maxima in free flow, a nearly circular ring with four maxima in the
coexistence region, and two in-file maxima above the transition.

\begin{figure}[t]
  \begin{center}
    \includegraphics[width=0.6\textwidth]{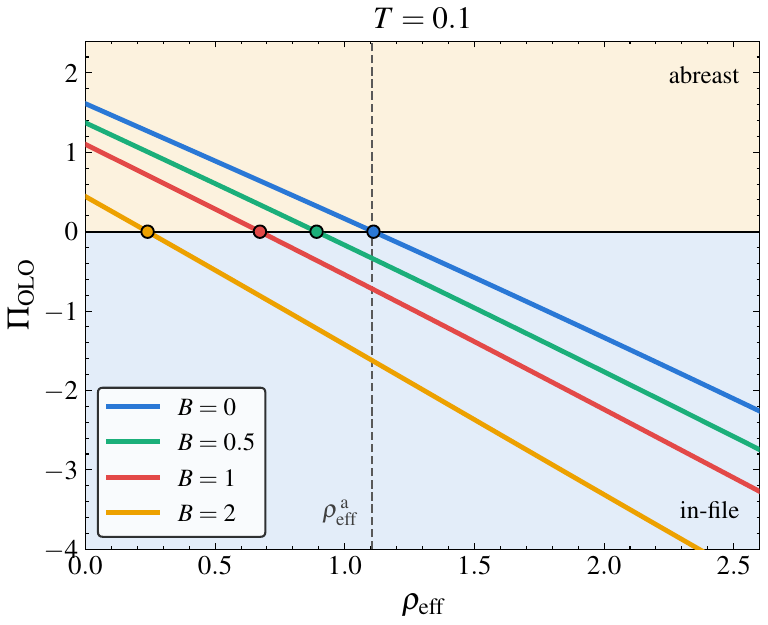}
    \caption{$\Pi_\text{OLO}$ of the model against $\rho_\text{eff}$ for four
      values of $B$, at $A=0.2$ and $T=0.1$. Dots mark the sign changes and
      the dashed line is $\rho_\text{eff}^{\,\text{a}}$. Positive values
      (amber) are abreast dominant, negative ones (blue) in-file, as in
      \fref{fig:olo-map}.}
    \label{fig:model-olo}
    \end{center}
\end{figure}

Using \eref{eqolo}, we compute $\Pi_\text{OLO}$ numerically for our model,
\eref{eqmodel}, integrating the Boltzmann weight of \eref{eqboltz} over the
two cones of \eref{eqcones} on a grid of $2400\times2401$ points in
$(u,\theta)$, with $u\in[0.03,9]$ (see \fref{fig:model-olo}). We observe that:

\begin{itemize}
\item $\Pi_\text{OLO}$ is close to linear in $\rho_\text{eff}$, as the
  measured $\Pi_\text{OLO}^\text{T}$ along its decreasing branch
  (\fref{fig:olo-map}(b)), and changes sign.
\item The density $\rho_\text{eff}^{\,\text{c}}$ of the sign change decreases
  as $B$ grows and, for $B>B_\text{c}$, lies inside the coexistence window,
  between $\rho_\text{eff}^{\,\text{f}}(B)$ and
  $\rho_\text{eff}^{\,\text{a}}$. The transition as the measurements define
  it, the density at which $\Pi_\text{OLO}$ changes sign, is therefore not a
  density at which the abreast or the in-file configuration gains or loses
  stability: the measured $\rho_c$ alone does not fix
  $\rho_\text{eff}^{\,\text{f}}$ or $\rho_\text{eff}^{\,\text{a}}$.
\item The temperature mostly sets the slope of the curve, i.e., at $B=1$,
  $\rho_\text{eff}^{\,\text{c}}=0.71$, $0.67$ and $0.60$ for $T=0.05$, $0.1$
  and $0.2$.
\end{itemize}

Around the transition, the cost is almost flat in $\theta$: at the density
$\rho_\text{eff}^{\,\text{c}}$, $W$ varies by $0.28\,T$ over the whole range of
$\theta$, against $1.2\,T$ in free flow and $2.1\,T$ at
$\rho_\text{eff}=1.8$. Close to the transition a dyad can therefore change
formation at almost no cost, e.g. to make room for an oncoming pedestrian.

\section{Discussion}
\label{sec:discussion}

A substantial fraction of the pedestrians in a crowd walk in small groups,
dyads above all, and modelling how these groups arrange themselves is a step
towards crowd models that describe the crowd as it is rather than as a
collection of individuals. Dyads walk abreast in a dilute crowd and in-file
in a dense one. This transition has only recently become measurable with high statistical accuracy, through a
real-life campaign of some of the authors that follows dyads across the full
range of naturally occurring crowd conditions \cite{vdLC2026}. Naturally,
since such data were not available before, pedestrian models in general, and
cost-function models of group formation in particular, have so far largely
neglected this transition, although realistic models should display it as an
emergent behaviour.

In this paper, focusing on dyads walking through a standing crowd, we have
proposed a cost-function model of dyad formation whose minima bifurcate in a
qualitatively correct way. As an effective density parameter increases, the
cost function passes from two abreast minima, to four minima, two abreast and
two in-file, whose sampling gives the ring-like distributions of the
transition, and finally to two in-file minima. Consistently with the
measurements, and differently from previous models, in the transition regime
the minima of our model (for parameters in the appropriate range,
$B>B_\text{c}$) lie only along the direction of motion of the dyad's centre of
mass and orthogonally to it, never at intermediate angles.

Together with the cost function, we have analysed its behaviour qualitatively
through bifurcation diagrams, which show under which conditions each
configuration, or ``phase'', of the dyad exists, and when two of them coexist.
Under a Boltzmann-like assumption we have mapped the cost function to a
probability density of relative positions, $\mathbb{P}(x,y)$. Through the
Orientation Log-Odds $\Pi_\text{OLO}$, which measures, in factors of two, how
much more likely the abreast configuration is than the in-file one, we have
shown that, as in the data, our model predicts a change of sign of
$\Pi_\text{OLO}$ at a critical density, although in the model $\Pi_\text{OLO}$
only decreases monotonically with it.

Our work is, however, only a first step towards quantitative cost-based models
of dyad formation. First, the non-monotonicity of the measured
$\Pi_\text{OLO}$ shows that the formation depends on the crowd density and
relative velocity in a complex way, even in the simple standing-crowd regime.
A quantitative model should recover this non-monotonicity, and would thus
need more bifurcation points, possibly controlled by more than one parameter.
Second, on the mathematical side, whenever the in-file configuration is stable
our model has a cuspidal minimum there. While this type of singularity is
possibly not a source of practical issues, it appears unphysical. Finally, we
have deliberately neglected the difference in preferred distance that dyads
show between the abreast and the in-file formation. We therefore leave the
development of quantitative models as a next challenge for the community.

\begin{acknowledgements}
  AC and CL wish to acknowledge the support of a starting grant by the Eindhoven Artificial Intelligence Systems Institute (EAISI), The Netherlands.
\end{acknowledgements}

\begin{ethics}
The research relies on pedestrian trajectory data collected via overhead sensors designed to ensure anonymity, with part of the dataset already publicly available. Consequently, no personally identifiable information or additional personal attributes were used or stored.
\end{ethics}

\begin{aiDeclaration}
A Large Language Model (Claude Opus 5.5) was used to revise the text and the figures of the manuscript. All included content has been checked and approved by the authors.
\end{aiDeclaration}

\begin{contributions}
The authors confirm contribution to the paper as follows: study conception and design: AC, CL, FZ; measurement analysis: CL; model and analysis: AC, FZ; manuscript drafting: AC, FZ.
 All authors reviewed the results and approved the final version of the manuscript.
\end{contributions}

\bibliographystyle{cdbibstyle} %
\bibliography{template} %

\end{document}